# Charged particle guiding and beam splitting with auto-ponderomotive potentials on a chip

Robert Zimmermann[1], Michael Seidling[1], Peter Hommelhoff[1]

[1]Department Physik, Friedrich-Alexander-Universität Erlangen-Nürnberg (FAU), Staudtstraße 1, 91058 Erlangen, Germany

*We report guiding and manipulation of charged particle beams by means of electrostatic optics based on a principle similar to the electrodynamic Paul trap. We use hundreds of electrodes fabricated on planar substrates and supplied with static voltages to create a ponderomotive potential for charged particles in motion. Shape and strength of the potential can be locally tailored by the electrodes' layout and the applied voltages, enabling the control of charged particle beams within precisely engineered effective potentials. We demonstrate guiding of electrons and ions for a large range of energies (from 20 to 5000 eV) and masses ($5 \cdot 10^{-4}$ to 131 atomic mass units) as well as electron beam splitting as a proof-of-concept for more complex beam manipulation. Simultaneous confinement of charged particles with different masses is possible, as well as guiding of electrons with energies in the keV regime, and the creation of highly customizable potential landscapes, which is all hard to impossible with conventional electrodynamic Paul traps.*

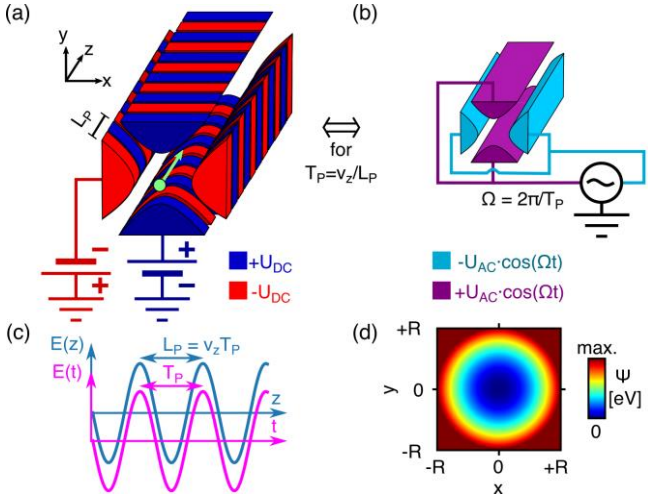

**FIG. 1. Principle of auto-ponderomotive guiding.** (a) When a beam of charged particles with velocity $v_z$ (green arrow) is injected into a structure consisting of electro*static* electrodes with spatially periodic voltages, the charged particle beam will be subjected to the equivalent transverse restoring force as charged particles in a linear electro*dynamic* trap with the same geometry but with unsegmented electrodes as shown in (b). This is because the spatially periodic electrostatic quadrupole field with period length $L_P$ leads to an alternating field with the periodicity $T_P = L_P/v_z$ in the rest frame of the moving particles (c). Like in a linear Paul trap, the charged particles experience a time-averaged harmonic pseudopotential, the ponderomotive potential, resulting in a restoring force towards the centerline. (d) The pseudopotential Ψ of both linear trap realizations shown in (a) and (b) can be made identical. R represents the electrodes' minimal distance from the guide's center.

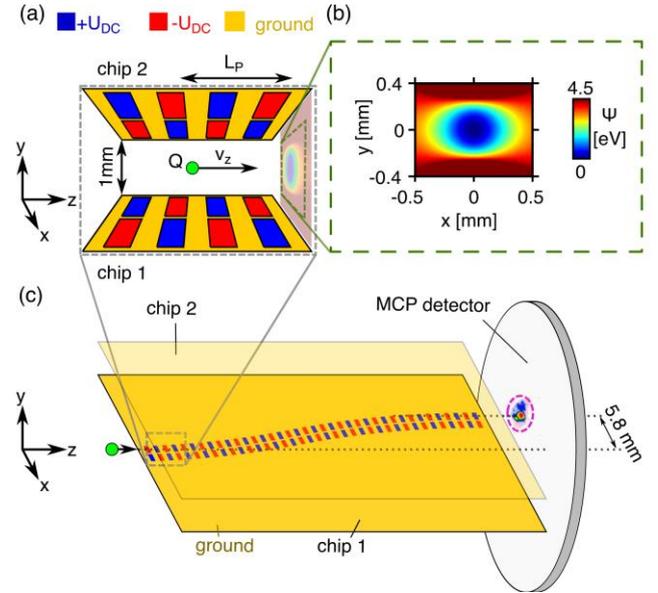

**FIG. 2. Auto-ponderomotive S-curved guide.** (a) A beam of charged particles with charge $Q$ and velocity $v_z$ is injected into a guiding structure consisting of two planar chips facing each other with a separation of 1 mm. The chips hold electrodes to which electrostatic potentials $+U_{DC}$ (blue) and $-U_{DC}$ (red) are applied. Their polarity varies periodically along the structure with the period length $L_P$ leading to the creation of the guiding pseudopotential for propagating electrons. (b) Simulation of the ponderomotive potential in a cut-plane transverse to the beam for electrons with $U_{DC} = 100$ V and $U_A = 1$ kV. The small ellipticity is due to the broken circular symmetry of the planar chips. (c) The electrodes on the chips define an S-curve that guides the particles so that they are laterally displaced. The particles are detected by a microchannel plate (MCP) detector 1 cm behind the structure. For illustration, only the bottom chip 1 is shown in full detail. Chip 2 has the mirrored electrode layout but with inverted polarity as shown in (a). The detector signal of guided particles is highlighted by a dashed purple circle. A picture of the front and back of the upper chip is displayed in the Supplementary Material in Fig. S2 on page 5.

The invention of radio frequency ion traps – Paul traps – 70 years ago set the foundation for precision mass spectrometry, ion trapping and cooling, and ion-based quantum computing [1-5]. These electrodynamic traps are based on the confining time-averaged forces exerted on charged particles by alternating electric fields [6]. In the following, we describe the electrostatic version of the electrodynamic Paul trap that dramatically expands the range of trapping parameters while maintaining the same operation principle. These devices can (unlike conventional electrodynamic Paul traps) simultaneously confine charged particles with vastly different masses in highly customizable potential landscapes. Importantly, the applicable electron energies for these new structures are high enough that they can be used in combination with a standard electron microscope as demonstrated in the electron beam splitting experiment below.

In our experiment, a charged particle beam with well-defined forward velocity is created and injected into a structure consisting of segmented electrodes with spatially alternating DC (direct current) voltages, as illustrated in Fig. 1. Like for magnetic undulators or strong focusing structures in particle accelerators [7], the electrostatic potential is transformed into an alternating potential in the rest frame of the moving particles and, thus, the particles are subjected to the same restoring transverse force as they are in a conventional linear Paul trap with AC voltages on non-segmented electrodes.

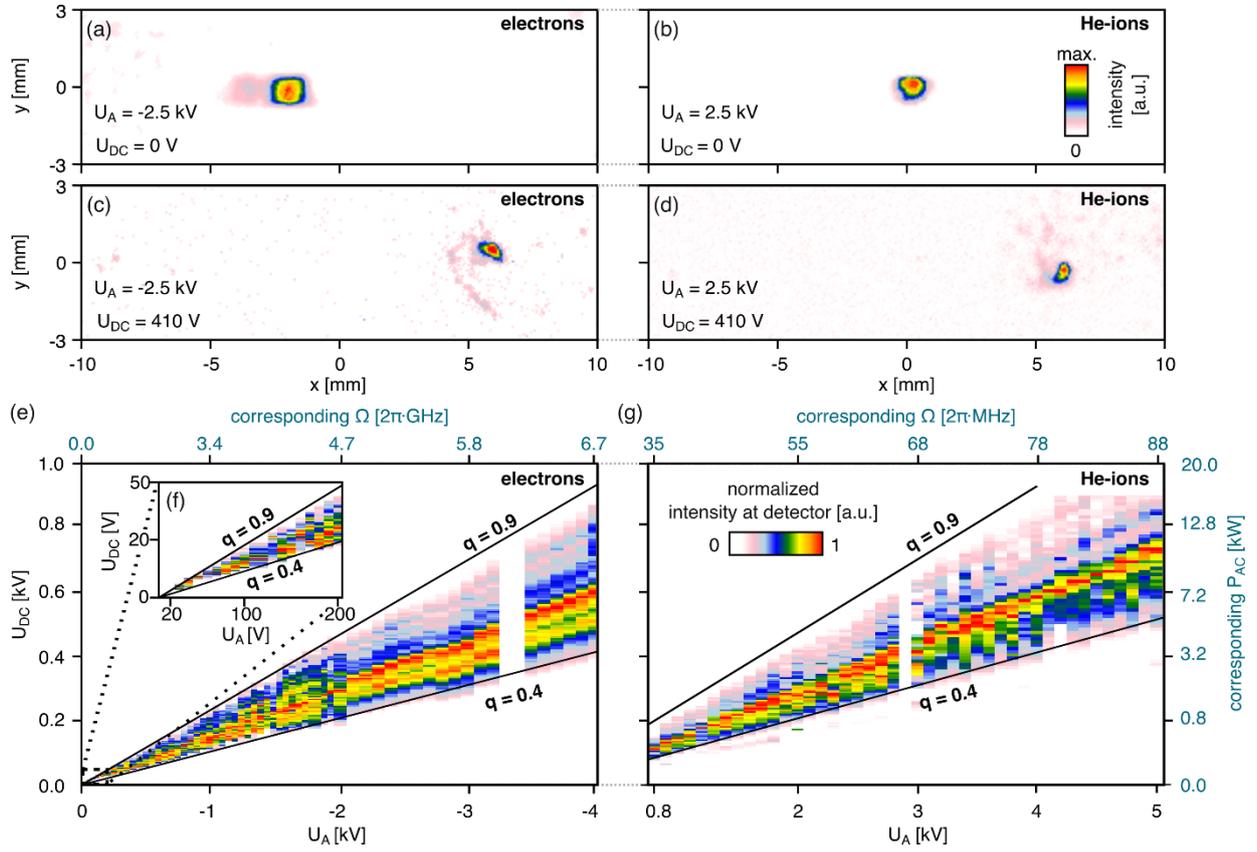

FIG. 3. **Auto-ponderomotive guiding of various species of charged particles.** (a), (b) Detector images of unguided electron (a) and helium ion (b) beams with electrodes grounded ($U_{DC} = 0$ V). (c), (d) With charged electrodes ($U_{DC} = 580$ V), both beams are guided and are measured at $x = 5.9$ mm (electrons) and $x = 6.1$ mm (He-ions), almost exactly at the expected position of $x = 5.8$ mm. The barely visible curl structure results from spiraling trajectories of off-centrally injected particles. The detector signal of helium ions is expected to be the vertically mirrored image of the electron signal, due to the opposite sign of their charge, which can just be discerned. (e), (g) Guiding stability: normalized intensity of guided electrons (e) and helium ions (g) on the MCP detector. For each acceleration voltage $U_A$, the applied electrode voltage $U_{DC}$ was scanned from 0-1 kV and the guiding signal of each scan was normalized to its maximum value. For comparison with electrodynamic traps, the corresponding driving frequency and AC power (impedance of 50 Ω) are given on the secondary axes (blue). Black lines corresponding to operation at $q = 0.4$ and $q = 0.9$ are drawn in (e) and (g) as a guide to the eye. Even though their masses differ by more than five orders of magnitude, guiding starts for all particles at $q = 0.4$ and no guiding is observed for $q$-values above 0.9, perfectly matching our particle tracking results. Because not all kinetic energies were possible to realize due to the source, some regions are left white. (f) Magnified image of a part of (e) showing that electrons are guided for kinetic energies as low as 20 eV, which was the lowest energy we could achieve with our source.

Since the driving frequency generating this ponderomotive force originates from the particles' forward velocity, we call the resulting effective potential "auto-ponderomotive". The resulting transverse forces are identical to those in a linear Paul trap. Hence, the stability of the trajectories of the charged particles is described by the two well-known dimensionless stability parameters $a$ and $q$ [1]. For non-relativistic particles, these parameters depend only on the amplitude of the applied voltages, the particle acceleration voltage $U_A$ and the guide's geometry. Yet, in stark contrast to electrodynamic traps, the stability parameters are independent of the charge-to-mass ratio here (the derivation is shown in the Supplementary Material on page 5). As a proof of concept, we present two auto-ponderomotive structures: One to show auto-ponderomotive guiding over a curved path and another one for auto-ponderomotive beam splitting.

Fig. 2 shows the design of a guiding structure and the simulation of the auto-ponderomotive potential $\Psi$. 84 electrodes are printed on each of the two chips and define an S-curve with a radius of curvature of $R_K = 0.535$ m, such that the output of the guide is laterally displaced by 5.8 mm with respect to its input. Static voltages $+U_{DC}$ (blue) and $-U_{DC}$ (red) are applied on the electrodes forming a system of quadrupole lenses with spatially periodic polarity (period length $L_P$=5.6 mm). This guide represents the electrostatic equivalent to the curved version of a conventional linear Paul trap with just an alternating potential applied ($a = 0$) [8]. The stability is therefore only determined by the parameter $q = \frac{\eta \cdot L_P^2 \cdot U_{DC}}{2\pi^2 \cdot R^2 \cdot |U_A|}$ (the derivation is shown in the Supplementary Material on page 5). The geometric factor $\eta = 0.61$ accounts for deviations from the ideal hyperbolic electrode geometry and from the perfectly sinusoidal form of the electric field [9]. $R = 0.5$ mm is the distance from the ponderomotive potential minimum to the chip surface.

The charged particle beams are unguided when the electrodes are grounded, as shown in the detector images for electrons in Fig. 3(a) and for helium ions in Fig. 3(b). When voltages are applied to the electrodes, guiding is observed for electrons [Fig. 3(c)], as well as helium ions [Fig. 3(d)], evidenced by a shifted detector signal at the expected guide exit position. In contrast to the unguided beams, the position of the guided beams on the detector remains unchanged even when magnets (~ 1 mT) are brought close to the vacuum chamber.

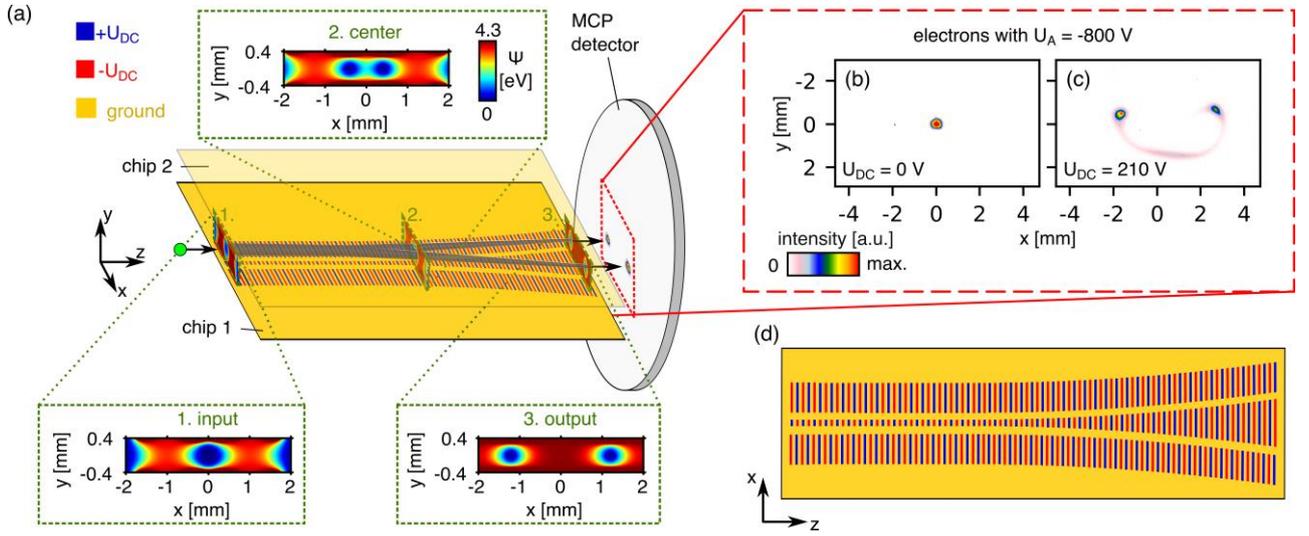

**FIG. 4. Auto-ponderomotive beam splitting on chip.** (a) The electrode layout is chosen such that the minimum of the ponderomotive potential is continuously split into two minima separated by 2.3 mm at the chip end. To illustrate this splitting, we plot the isopotential surface of $\Psi = 1$ eV for an electron beam with $U_A = 800$ V and $U_{DC} = 210$ V (transparent blue region). Simulated ponderomotive potential at the splitter's input (1), center (2) and output (3) are shown in the green dashed boxes. Clearly, the initially single central minimum splits continuously into two central minima as the particle propagates down the structure. The splitter consists of two chips facing each other with a separation of 1 mm. For illustration, only the lower chip is shown. The upper chip 2 has the same electrode layout but with inverted polarity. (b) Detector image of an unguided electron beam ($U_{DC} = 0$V). (c) Detector image of a split electron beam. Two spots are visible with a faint signal of lost electrons in between. The spot distance of 4.2 mm is expected given the opening angle of the split isopotential channels and the detector distance of 2.4 cm. (d) Top view of the beam splitter chip. The layout consists of three rows of electrodes. The width of the central electrodes widens along the chip from 0.3 to 2.2 mm (not to scale for illustration, see Fig. S3 in the Supplementary Material for a picture of the chip).

To characterize the guiding stability, we measure the guided number of particles for a range of particle beam energies and electrode voltages. For each $U_A$, the applied $U_{DC}$ was scanned from 0-1 kV. We observe guiding for a large energy range from 20 eV to 4000 eV for electrons [Fig. 3(e) and 3(f)] and from 800 eV to 5000 eV for helium ions [Fig. 3(g)]. Even though there are five orders of magnitude difference in the masses, guiding is observed for the same ratios of the applied voltages $U_{DC}/|U_A|$, corresponding to $q$ values between 0.4 and 0.9. The lower border in $q$ is due to the curvature of the guide. Since the restoring force of this potential depends on $q$, a finite value of $q \geq 0.39$ is needed to compensate the centrifugal force resulting from the curves (see the Supplementary Material), matching perfectly the experimentally observed minimum $q \cong 0.4$. The upper border corresponds similarly well to the maximum $q$ value of 0.91 in the first stability region of linear Paul traps [1]. The measurement was repeated for other noble gas ions (neon, argon, krypton, xenon) yielding similar results (displayed in the Supplementary Material in Fig. S4 on page 6). For comparison with radio frequency Paul traps, the corresponding driving frequencies and AC powers are given on the secondary axes in Fig. 3(c) and 3(e). Using electrons, the auto-ponderomotive design easily generates an apparent alternating field with driving frequencies in the gigahertz range with tens of kilowatts of AC power, which is virtually impossible to feed or maintain on an electrodynamic chip for thermal load reasons [8,10].

Next to this most versatile guiding demonstration, we now show that more complex potential landscapes can be realized based on the auto-ponderomotive principle. As an example, we show the design of an auto-ponderomotive beam splitter in Fig. 4. Here, the electrode layout of each chip consists of 270 electrodes forming three rows. Electrostatic voltages $+U_{DC}$ (blue) and $-U_{DC}$ (red) are applied, forming a system of multipolar lenses with spatially periodic polarity with a period length $L_P = 2.4$ mm. Moving along $z$, the width of the central electrodes widens (top view of the electrode layout is shown in Fig. 4(d)). This splits the initial central minimum of the ponderomotive potential into two minima, separated by 2.3 mm. A charged particle beam fed into the single central minimum at the structure input splits transversely into two beams following smoothly the auto-ponderomotive potential. This is shown in Fig. 4(c) for an electron beam (of a scanning electron microscope) as two distinct beam spots [unguided beam displayed in Fig. 4(b)]. A more detailed investigation of the beam splitter will be subject to forthcoming work.

The S-curved guide and the beam splitter presented here represent examples of what we call auto-ponderomotive engineering: A powerful method to create a custom-made ponderomotive potential landscape controlled by the geometry and arrangement of lithographically produced electrodes on a chip. It is noteworthy that the measured energy range for guided electrons from 20 to 4000 eV was limited by magnetic stray fields in the source for small beam energies and by sparking between the electrodes for large energies; we expect to expand the range from ~1 to at least 10 000 eV by shielding environmental magnetic fields and by using an optimized electrode layout for higher breakdown voltage. Because these structures work independently of the specific charge and mass of ions, any ion may be guided, hence simultaneous transport and trapping of various species of charged particles at well-defined velocities is possible, which might open up new possibilities to cold and collision chemistry [11-13]. Since the sensitivity of the charge-to-mass ratio in their electrodynamic counterparts is exchanged by a sensitivity on acceleration voltage, these structures work as energy filters when operated in analogy to a mass spectrometer [1]. Furthermore, we find from particle tracking simulations that the addition of (switchable) electron mirrors at both guide ends will allow the stable confinement of particles in three dimensions.

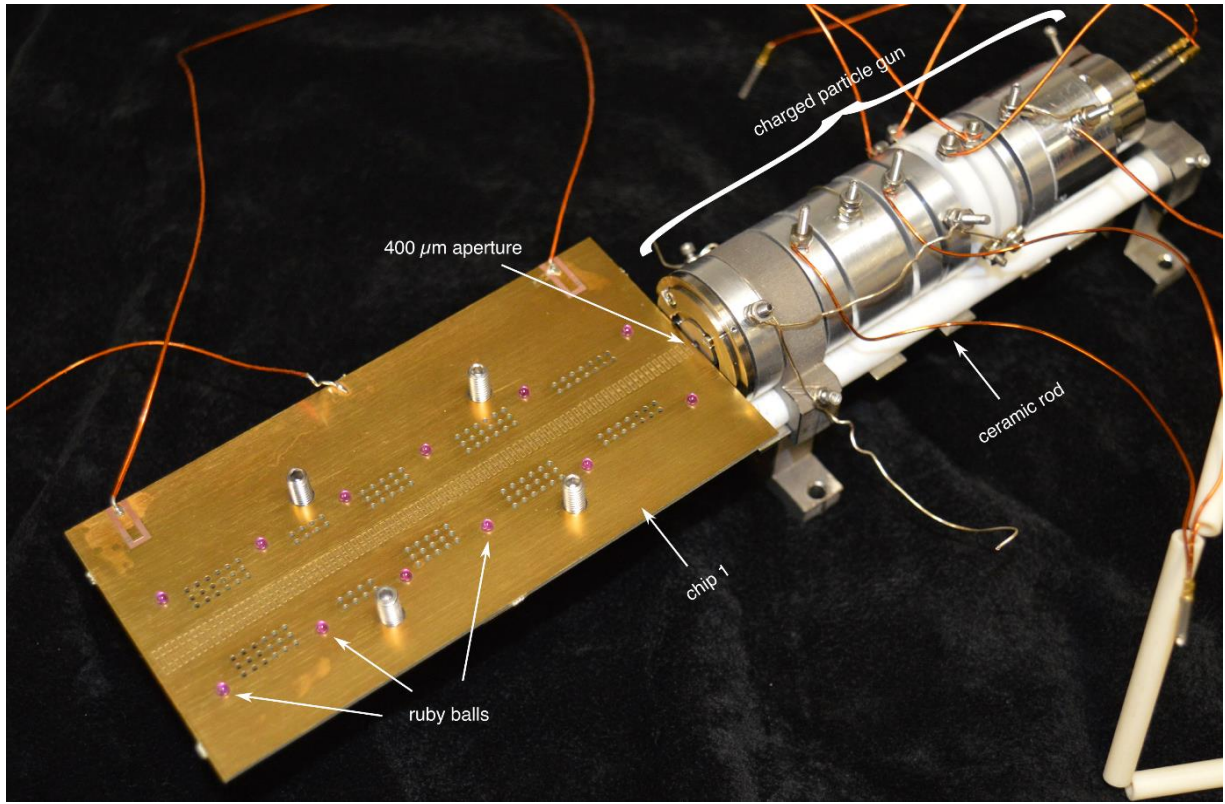

**FIG. S1. Picture of the electron-optical bench and charged particle gun.** The top chip has been removed for illustration purposes. All components are attached to the electron-optical bench with titanium clamps. The last aperture before the chip with a diameter of 400 µm limits the size and the divergence of the beam. The ruby balls on the chip ensure proper alignment of the two chips.

With the electrodes of the shown structures miniaturized to the micrometer scale, driving frequencies close to the terahertz ($10^{12}$ Hz) range can be achieved for electrons (see the Supplementary Material on page 7). This not only enters a new frequency range bridging the gap between microwave and optical frequencies, for which ponderomotive forces have been utilized [14-17], but it also facilitates extremely large trapping frequencies that may herald a new arena for quantum optics experiments and state-selective applications [18] with free electrons.

**Supplementary Material**

**1. Charged particle gun, electron-optical bench and detector**

A compact system consisting of a tungsten needle tip, extractor, four deflectors and two grounded apertures serves as a source of charged particles. The final aperture with a diameter of 400 µm limits the size and the divergence of the beam. The strong fields between the tungsten needle tip and the extractor allow electron field emission [19] as well as ionization of gas molecules [20], depending on the polarity of the applied acceleration voltage. Tips of varying sharpness and distance to extractor were used to realize particle beams of various energies. For the electron beam splitting experiment, a scanning electron microscope (Philips XL30 SEM), which illuminated a spot (<1 µm) at the input, is used as the electron source. All components are pre-aligned and fixed rigidly onto a 25 cm long electron-optical bench consisting of two straight ceramic rods. A picture of the setup is displayed in Fig. S1. The setup was placed into a vacuum chamber and the charged particles are detected with a microchannel plate (MCP) detector, which is placed 1 cm behind the S-curved guide (2.4 cm after the beam splitter). For the auto-ponderomotive guiding experiment, the measured intensity within a 1 mm wide square at point $x = 5.8$ mm and $y = 0$ mm on the detector is taken as the signal of the guided charged particles.

**2. Layout of the S-curved guide**

The S-curved guide consists of two planar chips facing each other with a separation of 1 mm. They are fabricated by a standard printed circuit board process on FR4 substrates with electrodes made from gold-plated copper. The chips have a total length of 11.3 cm and each chip consists of 84 electrodes. The electrodes define an S-curve with a radius of curvature of $R_K = 0.535$ m, such that the output of the guide is laterally displaced by 5.8 mm with respect to its input. The electrodes have a length of 1.3 mm and are 1.4 mm wide. The gap between the electrodes is 100 µm wide. The electrode layout of one of the two chips is displayed in Fig. 2 in the main text. The other chip has the mirrored electrode layout but with opposite polarity. Both chips have countersinks for ruby balls which serve to align the chips. The depth of the countersinks and the diameter of the ruby balls are chosen such that the chips are separated by 1 mm. The S-curved guide can be fixed and aligned to the electron-optical bench with a holder. A picture of the upper chip's front and back are shown in Fig. S2.

**3. Layout of the beam splitter**

The beam splitter consists of two planar chips facing each other with a separation of 1 mm. They are fabricated by a standard printed circuit board process on FR4 substrates with electrodes made from gold-plated copper. The chips have a total length of 11.3 cm and each chip consists of 270 electrodes arranged in three rows. The electrodes have a length of 0.55 mm and the gap between the electrodes is 50 µm wide.

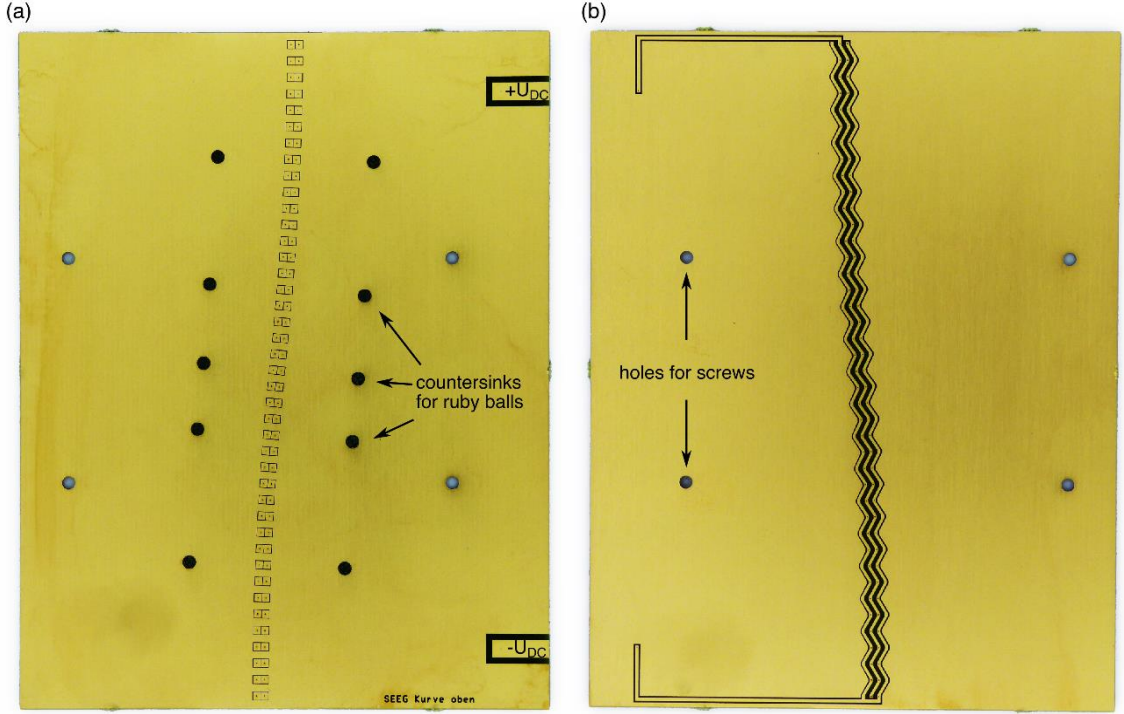

**FIG. S2. Front (a) and back (b) of the S-curved guide's upper chip.** Meandering electrodes on the back contact the electrodes on the front with plated through-holes (vias) resulting in spatially periodic voltages. The upper chip is placed above the lower chip such that their front sides are facing each other. Countersinks for ruby balls and holes for screws are drilled for alignment and fixation.

The outer electrodes have a width of 1.4 mm, while the width of the inner electrodes changes along the splitter from 0.3 mm to 2.2 mm. A picture of the electrode layout of one of the two chips is displayed in Fig. S3. The other chip has the mirrored electrode layout but with opposite polarity. Alignment and mounting of the chips are done as for the S-curved guides described above.

### 4. Derivation of the non-relativistic and auto-ponderomotive formula of the parameter $q$

The stability parameter $q$ for a linear Paul trap (with $\Phi_1 = U_{AC} \cos \Omega t$ and $\Phi_2 = -\Phi_1$ applied to adjacent rod electrodes) is given by $q = \frac{2Q \cdot 2 \cdot U_{AC}}{M \cdot R^2 \Omega^2}$ with $\frac{Q}{M}$ the charge-to-mass ratio of the charged particles, $R$ the minimal electrodes' distance to the guide center, $U_{AC}$ the amplitude of the alternating potential and $\Omega$ the driving angular frequency [1]. To derive the expression of $q$ for auto-ponderomotive guides, one replaces the driving frequency $\Omega$ with $2\pi \cdot \frac{v_z}{L_P}$. Here, $v_z$ is the velocity of the charged particles in the beam and $L_P$ the period length of the auto-ponderomotive structure. The velocity depends on the acceleration voltage $U_A$ as $v_z = \sqrt{2 \cdot QU_A/M}$. In our case, $U_{AC}$ must be replaced by $U_{DC}$. Hence, $q = \frac{L_P^2 \cdot U_{DC}}{2\pi^2 \cdot R^2 \cdot U_A}$. Because we use planar electrodes, a geometric factor $\eta$ needs to be included to describe the effective quadrupole strength of the used geometry [9]. Thus, the stability parameter is corrected to $q = \frac{\eta \cdot L_P^2 \cdot U_{DC}}{2\pi^2 \cdot R^2 \cdot U_A}$ and is valid for non-relativistic particles ($v_z \ll$ speed of light c). The sign of $q$ has no effect on the stability, therefore we only calculate the absolute value. An extension for relativistic velocities can be derived by including length contraction of $L_P$ and the Lorentz transformation of the electric field, but resulting expression is only independent of the particle's charge $Q$ and rest mass $M_0$ in the limiting cases of relativistically slow and fast particles.

### 5. Calculating $\eta$ and the harmonic region of the auto-ponderomotive potential

The ponderomotive potential is calculated as $\Psi = \frac{Q^2 \langle E^2 \rangle}{4 M \Omega^2}$ with $\langle E^2 \rangle$ the time-averaged squared electric field. For auto-ponderomotive structures, $\langle E^2 \rangle$ is calculated by the average of the electric field squared along the guide over the period length $L_P$. Compared to the ideal case of hyperbolic electrodes, the field strength of the quadrupole component is reduced by a geometric factor $\eta$ and is attained by a best fit from simulation. We obtain $\eta \approx 0.61$ for the guiding structure presented in Fig. 2 and Fig. 3 in the main text. Like in any harmonic approximation, the best fit is only valid close to the center. The discrepancy is less than 5% for displacements $\Delta x \leq 80$ μm from the guiding center and increases strongly for larger $\Delta x$.

### 6. Derivation of the minimum value of $q$ for the S-curved guiding structure

The harmonic force of the ponderomotive potential $F_H$ must compensate the centrifugal force $F_Z$ to guide the particles in a curve with curvature $R_K$. Stable trajectories are limited to the harmonic region of the guide, where the restoring force $F_H = -\nabla \Psi = -\nabla(\frac{1}{2}\omega^2 M \Delta x^2)$ reads $F_H = \omega^2 M \Delta x$ with the trapping frequency $\omega = \frac{q}{\sqrt{8}}\Omega$. The centrifugal force is given by the curvature $R_K$ and the particle's velocity $v_z$. Demanding that $F_H \geq F_Z$ leads to the guiding condition $q \geq \frac{L_P \cdot \sqrt{2}}{\pi \cdot \sqrt{R_K \cdot \Delta x}}$ resulting in a minimum value of $q = 0.39$ for a guiding structure with the geometry presented in this work, in excellent agreement with the experimentally observed $q_{min}$ for all guided species (see Fig. 3 in the main text and Fig. S4).

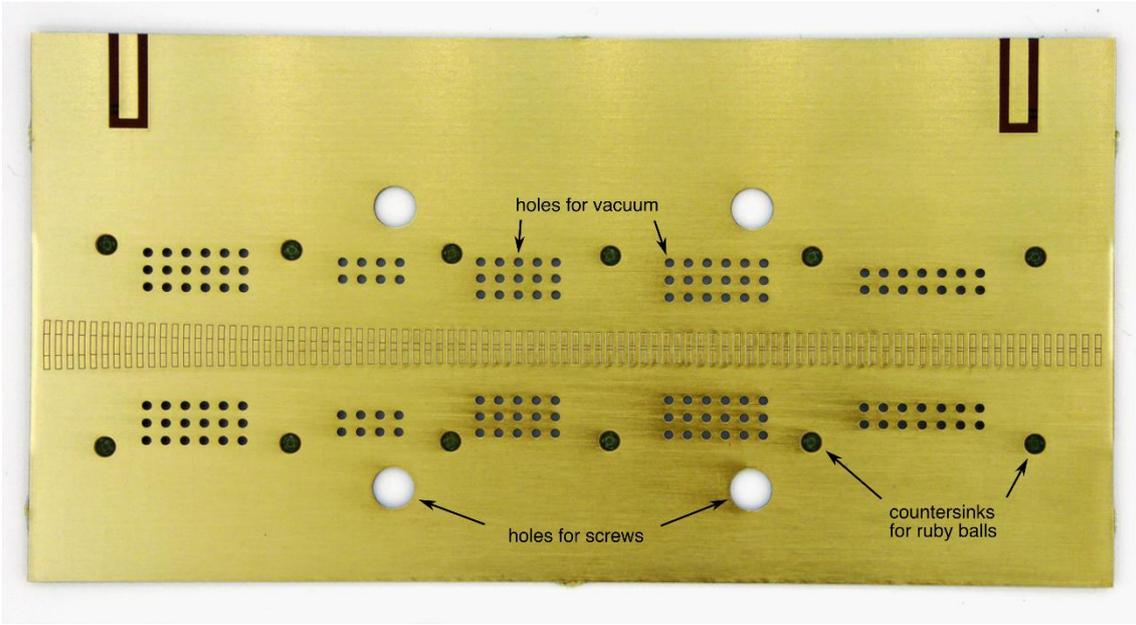

**FIG. S3. Front of the beam splitter's chip.** 270 electrodes arranged in three rows. The width of the inner electrodes changes along the splitter from 0.3 mm to 2.2 mm, while the width of the outer electrodes is 1.4 mm. Countersinks for ruby balls and holes for screws are drilled for alignment and fixation. Additional smaller holes are used to ensure good vacuum.

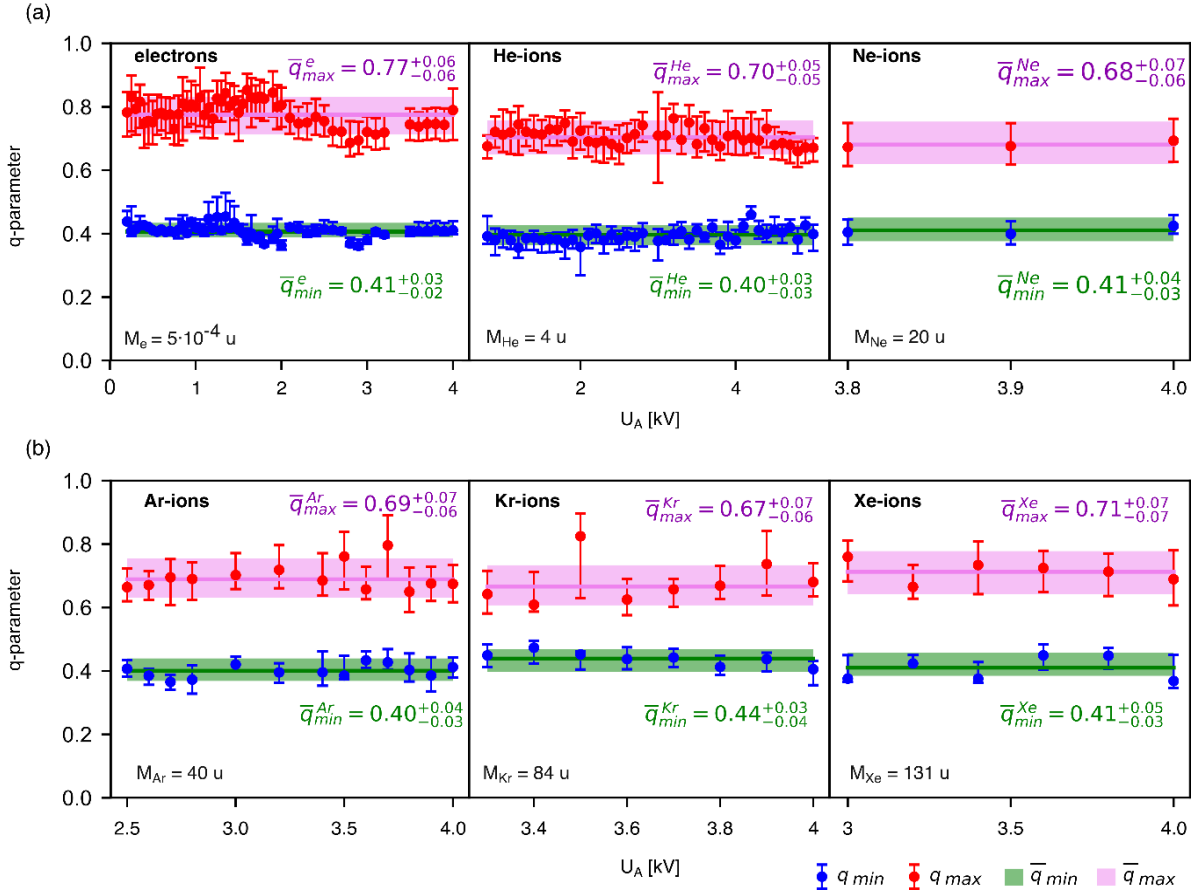

**FIG. S4. Auto-ponderomotive guiding of various species of charged particles.** (a), (b) The maximum and minimum $q$ values (defined as $q$ at 1/e the maximum intensity) for electrons and all used noble gas ions calculated from the detector intensity of the guided charged particles obtained by voltage scans (as discussed in the main text around Figure 3). As the guiding signal, the intensity within a 1 mm wide square at point $x = 5.8$ mm and $y = 0$ mm on the detector is taken. Note that the mass and charge-to-mass ratio varies over more than a factor of 200 000, clearly demonstrating that auto-ponderomotive potential engineering works independently of them. The measured maximum values of $q$ vary stronger than the minimum values because guiding becomes more sensitive to imperfect injection for high $q$ as the acceptance phase-space for coupling the beam into the guide decreases.

## 7. Miniaturization leads to higher trapping frequency ω

If the geometry of an auto-ponderomotive structure is scaled down by a factor of $c_g$, the period length $L_P$ and electrode's distance to the centerline $R$ are reduced to $L_P' = \frac{L_P}{c_g}$ and $R' = \frac{R}{c_g}$. The driving frequency $2\pi \cdot \frac{v_z}{L_p}$ increases accordingly to $\Omega' = \Omega \cdot c_g$. Since the stability parameter $q \propto L_P^2/R^2$ is independent of $c_g$, guiding is attained for the same applied voltage ratios for all scaling factors $c_g$ and the trapping frequency $\omega = \frac{q}{\sqrt{8}}\Omega$ increases to $\omega' = \omega \cdot c_g$. Thus, an electrode layout on the micrometer scale leads to much higher trapping frequencies if operated at the same stability parameter $q$. For example, using a guide with a period length $L_P$ of 56 µm ($c_g \sim 100$) (which is straightforward to manufacture) and an electron beam with a kinetic energy of 1 kV results in a driving frequency of $\Omega = 2\pi \cdot 0.33$ THz. Operating the guide at $q = 0.3$ ($U_{DC} = 77.4$ V, well below the breakdown voltage of high vacuum) leads to a trapping frequency of $\omega = 2\pi \cdot 36$ GHz.

## 8. Extended Data: Auto-ponderomotive guiding for electrons and noble gas ions

Fig. S4 displays the result for all used noble gas ions as discussed around Fig. 3 in the main text.